# COMPREHENSIVE UP-TO-DATE IMPACT OF THE IOMT IN HEALTHCARE AND PATIENTS

Guy Mouanda

Computer Science and Informatic Oakland University, Rochester MI

*ABSTRACT*

*The Internet of Medical Things (IoMT) is a quickly expanding field that intends to develop the features, effectiveness, and availability of healthcare services by applying numerous technologies to gather and diffuse medical data. IoMT devices incorporate wearable sensors, implantable devices, smart home methods, telemedicine policies, and mobile applications. IoMT applications range from chronic disease administration, remote patient monitoring, emergency response, and clinical decision support to health promotion and wellness. This paper aligns on the advantages, defies, and outlook directions of this developing domain. The paper also examines the ethical, legal, and social implications of IoMT, as well as the possible risks and vulnerabilities of the IoMT environment.*

## 1. INTRODUCTION

The Internet of Things (IoT) can be explained as a term that labels the network of physical objects implanted with sensors, and communication resources, granting them the ability to substitute data and proceed together with other devices, structures, and humans. IoT has been instructed in numerous domains, such as smart cities, smart agriculture, smart manufacturing, and smart transportation. He has exponentially grown to approximately 10 billion connected IoT devices at present, with a predicted increase to about 25 billion by 2025. [1] One of the most certifying and impactful fields of IoT is the Internet of Medical Things (IoMT), which examines the network of organized devices, sensors, and software that gather and communicate medical data. IoMT has the potential to reform the health care segment by granting advanced solutions for the diagnosis, treatment, anticipation, and management of numerous health conditions. IoMT can also improve patient involvement, decrease the health care costs, and advance health results and quality of life of individuals and populations.

According to a report by Grand View Research, the global IoMT market size was valued at USD 44.21 billion in 2023 and is expected to grow at a compound annual growth rate (CAGR) of 21.2% from 2024 to 2030.[2] The growth of IoMT is motivated by numerous aspects, such as the growing prevalence of chronic diseases, the mature population, the increasing request for remote healthcare services, the development of wireless technologies, the propagation of smart devices, and the rise of big data and artificial intelligence.

Therefore, IoMT is a complicated and multidisciplinary area that needs the partnership and management of diverse stakeholders, such as scientists, designers, contributors, regulators, and users. This paper aims to give a complete and up-to-date indication of the IoMT field, covering its primary applications, enabling technologies, the ethical, legal, and social implications of IoMT, and open issues. Furthermore, the paper recounts some smart future directions for IoMT research, such as edge computing, blockchain, artificial intelligence, and 6G. The paper is planned as follows: Section 2 presents the main applications of IoMT in healthcare; Section 3 describes the enabling technologies of IoMT; Section 4 discusses the open issues and challenges of IoMT; Section 5 talks about a benefit of IoMT, Section 6 highlights some future directions for IoMT research; and Section 7 concludes the paper and gives some recommendations

## 2. THE MAIN APPLICATIONS OF IoMT

In this section, we will investigate some of the primary applications of IoMT in healthcare, such as remote patient monitoring, chronic disease management, telemedicine, smart hospitals, and wearable devices. In the healthcare area, IoMT has many applications in which remote/self-health monitoring of numerous fundamental purposes such as heart rate, skin temperature, movement monitoring [3], monitoring of usual health conditions, nutrition status, and recovery of elderly or contaminated patients are more prominent to an improvement in life expectancy and a reduction in illness and death. [4]

### 2.1. Remote Patient Monitoring

Remote patient monitoring (RPM) is the use of IoMT devices to trace and transfer patients' vital signs, warning signs, and actions from their homes or supplementary spots to their healthcare providers. RPM can decrease the requirement for hospital appointments, reduce healthcare costs, and increase patient results and compensation. Examples of RPM tools are blood pressure checks, glucose meters, pulse oximeters, electrocardiograms, and smart scales. These tools can send out real-time data to the cloud, where it can be saved and investigated by healthcare specialists, and then run feedback, warnings, or involvements as required.

### 2.2. Chronic Disease Management

Chronic disease management (CDM) is the use of IoMT tools to support patients in their long-term health situations, such as diabetes, asthma, heart failure, or chronic obstructive pulmonary disease. CDM can enhance patients' faithfulness to cure proposes avoid barriers and improve their condition of life. Examples of CDM devices are insulin pumps, inhalers, medication dispensers, and smart patches. These devices can supervise patients' dose intake, symptoms, and biomarkers, and specify remembrances, notices, or modifications as needed. They can also associate with mobile apps or web portals, where patients can retrieve their health data, talk with their providers, and obtain training and help.

### 2.3. Telemedicine

Telemedicine is the use of IoMT devices to facilitate remote discussion, analysis, and healing between patients and healthcare providers. Telemedicine can improve the availability and cost-effectiveness of healthcare examination, specifically for rural, aging, or deprived populations. Examples of telemedicine devices are digital stethoscopes, otoscopes, thermometers, and cameras. These devices can obtain and transfer best-excellence audio and video data to the providers, who can then investigate, assess, and specify prescriptions or recommendations as demanded. They can likewise apply telemedicine platforms, such as video conferencing, messaging, or chatbots, to cooperate with the patients and give assistance and advice.

### 2.4. Smart Hospitals

Smart hospital technologies involve the use of IoMT devices to improve the surgical procedures, workflows, and ecosystems of healthcare areas. Smart hospitals can increase the security, effectiveness, and excellence of healthcare services, as well as the happiness and arrangement of patients and staff. Some examples of smart hospital devices are smart beds, infusion pumps, surgical robots, and ecological sensors. These devices can screen and modify the patients' environments, medications, and relaxation levels, as well as the hospital's warmth, lighting, and air circulation. They can also link with each other and with the hospital's information structures, such as electronic health records, record supervision, and security systems, to match up and plan responsibilities and actions. It is distinguished that by making the hospitals 'smart' equipment expenses might also be decreased due to the early discovery of irregularities that could distress the accuracy of exact interpretations from the medical devices, which could then lead to advanced maintenance expenses.[5]

## 2.5. Wearable Devices

The use of IoMT devices can be worn on the body, such as smartwatches, fitness trackers, smart glasses, and smart dressing, which are described as wearable devices. Individual health data like physical motion, heart rate, sleep quality, and stress levels can be gathered and evaluated with wearable devices. They can also facilitate users upgrading their health and wellness by granting feedback, proposals, or involvements. When a smart necklace notices a poor position it sends a notice to the user's smartphone, having a mobile app to fix his or her posture. [6] Wearable devices can also link to other IoMT devices, such as smartphones, tablets, or smart speakers, to improve their functionality and user capability.

## 3. ENABLING TECHNOLOGIES OF IOMT

In this section 3, we will sketchily propose some of the important supporting technologies of IoMT, such as wireless communication, cloud computing, edge computing, artificial intelligence, and blockchain.

## 3.1. Wireless Communication

Wireless communication is the spine of IoMT, as it allows the devices and sensors to connect with each other and with the cloud or edge servers. Wireless communication can be categorized into two categories: short-range and long-range. Short-range communication is a technology that enables communication within a local area network (LAN), such as Bluetooth, Wi-Fi, ZigBee, and Near Field Communication (NFC). These technologies are appropriate for low-power, low-cost, and low-data-rate applications, such as wearable devices, implantable devices, and smart home devices. Long-range communication authorizes communication over a wide area network (WAN), such as cellular networks, satellite networks, and low-power wide-area networks (LPWANs). These technologies are appropriate for high-data-rate, high-dependability, and high-exposure applications, such as remote monitoring, telemedicine, and smart city devices.

## 3.2. Cloud Computing

Cloud computing is a technology that assists in the integrated adaptation, treatment, and study of the data collected by IoMT devices and sensors. Cloud computing proposes various advantages for IoMT, such as scalability, flexibility, accessibility, and cost-efficiency. Cloud computing can also facilitate highly developed analytics and artificial intelligence functions, such as data mining, machine learning, deep learning, and natural language processing, that can obtain appreciated perceptions and information from the data. Nonetheless, cloud computing also has some restrictions for IoMT, such as inactivity, bandwidth, defense, and confidentiality. Cloud computing compels the data to be collected over the internet, which can establish delays, use system resources, and leak the data to possible attacks and violations. Consequently, cloud computing alone may not be enough for some IoMT applications that demand real-time, low-latency, and secure data management and decision-making.

### 3.3. Edge Computing

Edge computing is the technology that grants the storage, management, and investigation of the data at the periphery of the network, near the source of the data. Edge computing can match cloud computing by focusing on some of its restrictions for IoMT. Edge computing can decrease the potential, bandwidth, and security dangers of data communication, as well as facilitate local and context-aware data handling and assessment. Edge computing can also augment the consistency, flexibility, and independence of the network, as it can work separately from the cloud in case of network collapses or troubles. Edge computing can be applied by numerous devices and stages, such as smartphones, tablets, laptops, gateways, routers, and servers, that can act as edge nodes and deliver the essential computation, storage, and communication competencies for the IoMT devices and sensors.

### 3.4. Artificial Intelligence

Artificial intelligence is the technology that presents the intelligence, learning, and reasoning proficiencies for the IoMT devices, sensors, and functions. Artificial intelligence can be used in numerous purposes and structures for IoMT, such as data combination, data density, data quality valuation, data anomaly detection, data organization, data forecast, data visualization, data reference, data feedback, and data action. Artificial intelligence can also assist in various treatments and services for IoMT, such as diagnosis, prognosis, treatment, prevention, wellness, education, and investigation. Artificial intelligence can be functional at dissimilar stages of the IoMT network, such as the device level, the edge level, and the cloud level, subject to data accessibility, data difficulty, data privacy, and data expectancy constraints. AI can enable discovery, large-scale screening, monitoring, resource provision, and prediction of potential interactions with the new anticipated therapies. [7]

### 3.5. Blockchain

Blockchain is the technology that supports the dispersed record, agreement, and cryptography methods for the IoMT network. Blockchain can recommend some advantages for IoMT, such as security, privacy, faith, clarity, and liability. Blockchain can guarantee the confidence and discretion of the data by encrypting it and saving it in a distributed and permanent record that is allocated and proved by the network users. Blockchain can also warrant the confidence and clarity of the information by presenting a visible and auditable record of the data dealings and procedures that appear in the network. Each record contains a cryptographic hash of the previous records to chain the records together and make them resistant to modifications.[8] Blockchain can also make sure the liability of the data by allowing smart agreements and inspirations that can inflict the rules, policies, and contracts of the data revealing and control between the members of structures.

## 4. ISSUES AND CHALLENGES

This section presents the risks, vulnerabilities, challenges, and issues of the Internet of Medical Things. IoMT poses significant challenges that need to be addressed, such as data security, privacy, regulation, and ethical issues.

Health Informatics - An International Journal (HIIJ) Vol.13, No.3, August 2024

- Data security and privacy: IoMT mechanisms create and diffuse significant quantities of complex health information, which can be susceptible to cyberattacks, data violations, unlawful approaches, and exploitation. IoMT devices also increase privacy worries, as patients may not have maximum control or knowledge of how their data is stored, shared, and used, and by whom. IoMT devices may also cause threats of identity theft, fraud, discrimination, and abuse. Due to the sympathetic nature of the information it contains about a person's health, it is very vital to keep healthcare data protected and stop unapproved people from gaining access to your network. [9]
- Data consistency: IoMT devices frequently run on diverse policies, rules, and designs, which can delay the incorporation and interchange of information between diverse devices, coordination, and stakeholders. IoMT devices also lack conventional principles and regulations for data excellence, authenticity, consistency, and equivalence. This can impact the truth, entirety, and usability of the data and limit the capabilities of data analytics and artificial intelligence.
- Data regulation and governance: IoMT mechanisms are exposed to dense, and changing governing and lawful frameworks, which alter across nations and regions. IoMT devices also increase
- Moral and social concerns, such as data proprietorship, permission, responsibility, precision, dependence, and justice. IoMT devices complement and provide steady guidelines and methods for data gathering, storage, communicating, and use, as well as for data safety, accordance, and management.
- Accountability and liability: Internet of Medical Things tools encounter complex accountability and liability situations and involve relationships between humans and machines, which can establish uncertainty and doubt about who is accountable and liable for the significance of the devices. For example, if an IoMT device crashes or fails due to a design fault, a software bug, a human error, a cyberattack, or a natural disaster, who should be held responsible and accountable for the losses or damages produced by the device? How can causality and fault be decided and developed in such events? How can the rights and advantages of the users and the sufferers be covered and rewarded?
- Equity and justice: IoMT devices can establish new prospects and benefits for some groups of people, but also new dangers and damages for others. For example, IoMT devices can increase the approach and affordability of healthcare for people in remote or underserved zones, but they can also aggravate the digital divide and the health inequalities for people who lack the income or knowledge to use the devices. IoMT devices can likewise facilitate identified and protective healthcare for people, but they can also create additional methods of prejudice and humiliation based on the information gathered and evaluated by the devices. How can the advantages and damages of IoMT devices be spread honestly and rightly among different groups of people? How can the beliefs and favorites of users and society be displayed and respected in the design and distribution of IoMT devices?

These are a portion of the major moral, legal, and social ramifications of IoMT that involve significance and reflection by all partners engaged with the turn of events, guidelines, and utilization of IoMT gadgets. The challenges and prospects in the IoMT field are continually shifting and increasing. Akhbarifar et al. (2020) explained a lightweight block encryption technique for remote health monitoring with conditions for the security of health and medical data in a cloud-based IoMT setting. [10]

The field is still developing. Consequently, it is vital to encourage a cooperative and multidisciplinary way to deal with and address the moral, legitimate, and social issues of IoMT and guarantee that IoMT gadgets are planned and utilized in a capable, reliable, and useful way for the individual and the public.

## 5. BENEFITS OF IoMT

This section provides a brief overview of IoMT benefits. Developed patient commitment and authorization: IoMT gadgets can permit patients to watch their personal health specifications, read their health histories, correspond with their health care providers, and contribute to their personal care choices. IoMT gadgets can also plan feedback, learning, and incentives for patients to follow their medicine proposals and implement healthy behaviors. Many IoMT improvements, rights, and healthcare resolutions may probably benefit healthcare routines [11]. IoMT can suggest countless advantages for health care, such as:

- Developed excellence and effectiveness of care: IoMT gadgets can grant real-time, true, and inclusive information on patients' health significance, warning signs, medicine obedience, and response to interferences. This can facilitate health care providers to detect, handle, and block diseases, as well as to raise care paths and resource provision. IoMT devices can also enable remote care release, dropping the necessity for hospital appointments and charges and refining admission and accessibility for patients and providers.
- Decreased health care budgets and waste: IoMT devices can decrease healthcare budgets by permitting immediate revealing and involvement, dropping hospitalizations and readmissions, developing care organizations, and decreasing mistakes and repetition. IoMT gadgets can also facilitate lower healthcare waste by reducing the use of unnecessary analyses, trials, and medicines and by adjusting the application of healthcare tools and infrastructure.

## 6. FUTURE DIRECTIONS OF IoMT

This section highlights the future direction of the IoMT and presents some essential issues and allegations for healthcare stakeholders. IoMT is still growing and confronts indecision about its future directions and waves, IoMT is still a promising and vibrant field, that is persuaded by numerous issues, such as technological improvement, marketplace challenges, consumer inclinations, health care needs, and governing and guidelines improvements. Some of the viable future directions of IoMT are:

- Enhanced acceptance and diffusion of IoMT gadgets: IoMT devices are presumed to be more extensive and varied, as extra health care providers, payers, patients, and users accept and use them for several objectives and venues. IoMT gadgets are also estimated to develop additional reasonable, available, and user-friendly, option as well as more included and interoperable devices and systems.
- Improved functionality and intelligence of IoMT gadgets: IoMT devices are predicted to become more innovative and classier as they integrate new structures and abilities, such as biometrics, biosensors, nanotechnology, robotics, 3D printing, and amplified and virtual reality. IoMT gadgets are also believed to become smarter and more independent, as they control information analytics and artificial intelligence to grant extra personalized, analytical, and regulatory health care solutions.

- Developed capacity and power of IoMT devices: IoMT devices are assumed to have expansive and meaningful effects on healthcare as they facilitate new prototypes and styles of care distribution, such as correctness medicine, population health, and value-based care. IoMT devices are similarly anticipated to have broader and deeper inplications for health care, as they touch on the roles and relations of healthcare stakeholders, the quality and products of health care, and the health and well-being of individuals and groups.

Another future direction of IoMT is to implement a security-by-proposal philosophy and models for IoMT gadgets, such as encryption, authentication, and patching. Apply strong guiding principles and procedures for data security and privacy, such as educated permission, data minimization, and anonymization. Obtain successful power and oversee processes for IoMT devices, such as certification, auditing, and reporting. Instruct and teach consumers and workers about the advantages and threats of IoMT gadgets, as well as the best methods for their use and protection. Final steering; nonstop investigation and assessment of influence and consequences of IoMT devices, as well as the developing risks and encounters.

The sixth generation of wireless technology, or 6G, is predicted to be available in the 2030s, and it will pose outstanding abilities for IoMT. 6G will facilitate ultra-high-speed, ultra-low-latency, ultra-reliable, and ultra-massive interaction, as well as improved aspects such as artificial intelligence, holography, and quantum computing. 6G will also strengthen new models of networking, such as terahertz, orbital, and underwater communication, as well as their combination with satellite and drone techniques. 6G and IoMT will transform healthcare in the future, by allowing new and better-quality applications and situations that will improve the quality, competence, and convenience of healthcare services, as well as the authorization and well-being of patients and providers.

## 7. CONCLUSION AND RECOMMENDATIONS

IoMT is a quickly changing and growing field that suggests excellent prospects and argues for health care and humanity. IoMT gadgets and applications can raise the supremacy, availability, and effectiveness of health care, as well as persuade patients and health care providers with supplementary, identified, and appropriate data and involvements. Conversely, IoMT furthermore proposes substantial ethical, legal, and social encounters that demand to be talked about by several stakeholders, such as healthcare professionals, patients, regulators, manufacturers, and researchers. These challenges comprise warranting the safety, security, and dependability of IoMT devices and applications; protecting the privacy, confidentiality, and agreement of health data; ensuring the fairness, accountability, and transparency of IoMT algorithms and decisions; appreciating the independence, self-respect, and favorites of patients and health care providers; and matching the profits and dangers of IoMT for individuals and society. IoMT needs a multidisciplinary and cooperative style that includes ethical, legal, and social analysis, valuations, and governance, as well as public engagement and teaching, to warrant that IoMT is advanced and used in an accountable, beneficial, and supportable way. One way to improve the quality of care given to patients and their general health is to employ the handful of pieces of medical technology and applications that have been barely built for the medical field. [12]
Some of the suggestions for future research and advancement of the IoMT are:

- Extend and approve mutual specifications, procedures, and structures for the IoMT to ensure data interoperability, compatibility, and quality.
- Employ and impose strong security and privacy rules for the IoMT to defend data confidentiality, integrity, and availability.

- Construct and estimate user-centric and human-centric IoMT results to guarantee user belief, compensation, and confidence.
- Investigate and direct the moral, social, and legal consequences of the IoMT to confirm user self-respect, independence, and integrity.
- Cooperate and align with several participants, such as patients, providers, researchers, developers, regulators, and policymakers, to guarantee the configuration of the IoMT with the requirements, prospects, and beliefs of the society.